\documentclass[a4paper,aps,pra,floatfix,tightenlines,superscriptaddress,onecolumn,nofootinbib]{revtex4}
\usepackage[utf8x]{inputenc}
\usepackage{graphicx}
\usepackage{amsmath}
\usepackage{amsthm}
\usepackage{amsfonts}
\usepackage{amssymb}
\usepackage{paralist}
\usepackage{bbm}
\usepackage{latexsym}
\usepackage{multirow}
\usepackage[bookmarks = true, colorlinks = true, urlcolor=blue]{hyperref}



\begin{document}

\title{Optomechanical Schr\"odinger Cats -- a Case for Space}
\author{Rainer Kaltenbaek}
\affiliation{Vienna Center for Quantum Science and Technology (VCQ), Faculty of
Physics, University of Vienna, Boltzmanngasse 5, A-1090 Vienna, Austria}
\email[Corresponding author: ]{rainer.kaltenbaek@univie.ac.at}

\author{Markus Aspelmeyer}
\affiliation{Vienna Center for Quantum Science and Technology (VCQ), Faculty of
Physics, University of Vienna, Boltzmanngasse 5, A-1090 Vienna, Austria}

\begin{abstract}
\noindent Quantum optomechanics exploits radiation pressure effects inside optical cavities. It can be used to generate quantum states of the center-of-mass motion of massive mechanical objects, thereby opening up a new parameter regime for macroscopic quantum experiments. The challenging experimental conditions to maintain and observe quantum coherence for increasingly large objects may require a space environment rather than an earth-bound laboratory. We introduce a possible space experiment to study the wave-packet expansion of massive objects. This forms the basis for Schr\"odinger cat states of unprecedented size and mass.
\end{abstract}

\maketitle

The transition from the classical world view to the one of quantum physics was neither smooth nor fast -- in fact it may still be going on. In particular, the first decades of the 20th century saw the emergence of a multitude of new concepts -- introduced in order to understand experimentally observed effects that could not be grasped with the tools of classical physics: matter waves, probability amplitudes, complementarity and uncertainty relations come to one's mind immediately, just to name a few examples. And not even when quantum theory had taken concise forms did the debates about the fundamental relevance of its underlying concepts end.

Einstein and Schr\"odinger, for example, took both very different viewpoints, which again differed from Bohr's~\cite{Howard2004}. The famous work of Einstein, Podolsky and Rosen (EPR) criticizes the fact that quantum physics provides an incomplete description of what EPR define as ``physical reality''~\cite{Einstein1935}, i.e., the notion that, prior to observation, all physical properties are in some sense definite. Bohr responded that this incompleteness is an intrinsic feature linked to the fact that no experimental situation can consistently define all possible elements of physical reality~\cite{Bohr1935}~\footnote{``In fact, it is only the mutual exclusion of any two experimental procedures, permitting the unambiguous definition of complementary physical quantities, which provides room for new physical laws, the coexistence of which might at first sight appear irreconcilable with the basic principles of science.''} and hence no more complete description of physical reality will be possible in principle. In a way, Bohr's response anticipated the famous results of Bell~\cite{Bell1964}, Kochen and Specker~\cite{Kochen1968}, and later Greenberger, Horne and Zeilinger~\cite{Greenberger1990}. Curiously, in a situation similar to the one discussed by EPR, Schr\"odinger takes the opposite standpoint to Einstein and concludes his analysis by issuing ``an emergency decree: in quantum physics statements about what 'really' is, statements about the object, are forbidden, they only deal with the object-subject relation''\footnote{``\ldots'Notverordnung': in der Quantenmechanik sind Aussagen \"uber das, was 'wirklich' ist, Aussagen \"uber das Objekt, verboten, sie handeln nur von der Relation Objekt-Subjekt \ldots''(Letter from Schr\"odinger to Sommerfeld; December 11 1931; see \cite{vonMeyenn2011a})}.

Both Schr\"odinger and Einstein were of the opinion that one has to go beyond ``dogmatic quantum mechanics''~\footnote{``Ich hab mich sehr gefreut, da\ss{} Du in der eben erschienen Arbeit im Physical Review die dogmatische Quantenmechanik bei dem Schlafittchen erwischt hast...'' (Letter from Schr\"odinger to Einstein; June 7 1935; see \cite{vonMeyenn2011a}).} Yet in their approaches, to quote Einstein, ``they were the strongest antipodes''~\footnote{``Dabei sind wir in der Auffassung des zu erwartendenen Weges sch\"arfste Gegens\"atze'' (Letter from Einstein to Schr\"odinger, August 8 1935; see \cite{vonMeyenn2011a}).}. 

One well-documented example (of great impact) is the question of how to interpret the quantum-mechanical wavefunction. Einstein believed that one can only meaningfully talk about the wavefunction in an ensemble sense, i.e., there must be a statistical character to the wavefunction ~\footnote{``\ldots die $\Psi$-Funktion beschreibt nicht einen Zustand eines Systems sondern (statistisch) ein Ensemble von Systemen. \ldots Nat\"urlich l\"asst diese Deutung der Quantenmechanik besonders klar hervortreten, da\ss{} es sich um eine durch Beschr\"ankung auf statistische Aussagen erkaufte M\"oglichkeit unvollst\"andiger Darstellung der wirklichen Zust\"ande und Vorg\"ange handelt.'' (Letter from Einstein to Schr\"odinger; August 8 1935; see \cite{vonMeyenn2011a}).}. Schr\"odinger however, at least in the early 1930s, believed that the wavefunction resembles some sort of physical reality~\footnote{``Du aber siehst als Ursache der inneren Schwierigkeiten etwas ganz anderes. Du siehst in $\Psi$ die Darstellung des Wirklichen ..." (Letter from Einstein to Schr\"odinger; August 8 1935; see \cite{vonMeyenn2011a}).}. Their discussions eventually resulted in the formulation of at least two variants of a gedankenexperiment that involed the superposition of two macroscopically distinct states of a physical system -- the states ``exploded'' and ``not exploded'' of a pile of gunpowder (Letter from Einstein to Schr\"odinger; August 8 1935; see \cite{vonMeyenn2011a}) and the notorious cat states ``dead'' and ``alive''~\cite{Schroedinger1935}. Even today, this gedankenexperiment captures one of the most outstanding questions on the conceptual foundations of quantum physics. Are the laws of quantum physics universally valid? If so, what does that mean for the physical reality of macroscopically distinct states? Beyond being of significant philosophical relevance, these questions pose a very concrete experimental challenge. 

Whether it is possible to create superpositions of macroscopically distinct states of arbitrary distinctness has not yet been decisively answered. So far, all quantum experiments, even those involving some level of macroscopicity, are consistent with the predictions of quantum theory~\cite{Leggett2002a,Nimmrichter2013a}. Some of the most prominent examples include the delocalization and interference of macromolecules~\cite{Arndt1999} of up to 7,000 atomic mass units (amu)~\cite{Gerlich2011}; the coherent superposition of co- and counterrotating macrocopic currents of some $\mu$-Ampere~\cite{Friedman2000,VanderWal2000}; the generation and decoherence of Schr\"odinger cat states of photons~\cite{Deleglise2008,Hofheinz2009} and atoms~\cite{Monroe1996,Myatt2000,Haffner2005,Leibfried2005}; the generation of spin-entanglement between distant ensembles of up to $10^{12}$ atoms~\cite{Julsgaard2001} and between different spatial modes of a Bose-Einstein condensate (BEC)~\cite{Gross2011,Li2010}; the generation of entanglement between photons that differ by $600$ in their orbital angular momentum quantum number~\cite{Fickler2012} or of superpositions of two spatially different (energy eigen-) states of motion of a micromechanical resonator~\cite{O'Connell2010}. Pushing the parameter range of such experiments even further remains an intriguing challenge of experimental quantum science. Here, we introduce a new approach based on the quantum optical manipulation of the center-of-mass motion of macroscopic objects, i.e. quantum optomechanics~\cite{Aspelmeyer2012}. In principle, this scheme allows for a significant extension of the parameter regime of Schr\"odinger cats with respect to both size and large spatial distinctions in the superposition of different center-of-mass states. Given sufficiently large masses, such spatial optomechanical Schr\"odinger cats could even allow to explore the interface between quantum physics and gravity.

The young field of quantum optomechanics exploits the methods of quantum optics to prepare, manipulate and analyze nano-, micro- and macro-mechanical devices in the quantum regime of motion. The main idea, which can be traced back to the early days of gravitational wave interferometers~\cite{Braginsky2001a,Braginsky2002}, is to make use of radiation-pressure interactions between the mechanical object and the radiation field of a high-quality optical cavity. In essence, when the mechanical motion can modify the cavity response, e.g., via direct modulation of the cavity length or via dispersion, the resulting radiation-pressure force both depends on the (center-of-mass) position degree of freedom and, because of the finite cavity lifetime, is retarded in time. This gives optical control over the full mechanical susceptibility, i.e., over both spring constant and damping of the mechanical resonator. For sufficiently strong forces and specifically chosen optical states, this allows full (quantum) optical control over quantum states of massive mechanical oscillators. 

For example, analogous to laser cooling of atoms and ions~\cite{Diedrich1989,Hechenblaikner1998,Vuletic2000a,Wilson-Rae2004}, quantum optomechanics allows for cooling the motion of mechanical devices~\cite{Gigan2006,Arcizet2006a,Kleckner2006a,Schliesser2006,Wilson-Rae2007,Marquardt2007,Genes2008} and, only recently, several demonstrations of laser-cooling nano- and micromechanical resonators into the quantum ground state of motion have been reported~\cite{Teufel2011,Chan2011c}. In parallel, experiments are now entering the so-called strong coupling regime~\cite{Brennecke2008,Murch2008,Groblacher2009a,O'Connell2010,Teufel2011a,Verhagen}, which allows overcoming effects of decoherence and enabling coherent interactions between photons (of the cavity field) and phonons (of the mechanical resonator). 

The following discussion is strongly motivated by realizing that quantum optomechanics offers an almost universal coupling mechanism for controlling the center-of-mass degree of freedom of massive objects -- independent of their size. Its working principles have now been demonstrated for mechanical objects spanning a mass range of almost twenty (!) orders of magnitude in mass: from clouds of up to $10^7$ ultracold atoms with $10^{-20}$ kg over nano- and micromechanical resonators of up to $10^{14}$ atoms with $10^{-10}$ kg to massive mirrors of gravitational wave detectors with more than $10^{20}$ atoms and a weight of several kg~\cite{Aspelmeyer2012}. Obviously, for any of the chosen objects, entering the quantum regime will require a close look at the possible sources of decoherence.

Typical mechanical resonators are rigidly clamped to a support structure, which comes with several disadvantages: first, it resembles a permanent strong link to a thermal environment; second, it limits the maximum displacement achievable for an optomechanical cat state; and third, internal loss mechanisms can result in highly non-Markovian Brownian motion~\cite{Groblacher2013a}. Recently, there have been several proposals how to strongly suppress these effects in a quantum-optomechanics configuration by using optically trapped nano- or microspheres \cite{Chang2010a, RomeroIsart2010a, Barker2010a}. In analogy to experiments with trapped atoms and ions, the optical trap provides a harmonic potential for the center-of-mass motion of the nano- or microsphere that now represents the mechanical oscillator. This configuration forms the basis for our analysis. 

We consider a situation in which we can cool the center-of-mass of a nano- or microsphere close to its quantum ground state of motion (see \cite{Kaltenbaek2012b} for more details on the experimental requirements). After successful cooling, the particle will be released and its wave-packet will expand rapidly. By measuring the dynamics of this wave-packet expansion, i.e., by measuring the size of the wave-packet as a function of time, it is possible to test the predictions of quantum theory and to analyze the precise nature of the decoherence mechanisms affecting the quantum state. This is because, as the size of the wave-packet increases, the particle will be in a quantum superposition of being anywhere within an expanding region of space. As a consequence, any additional decoherence, for example resulting from collisions with gas particles or from scattering, absorption or emission of blackbody radiation, will not only lead to localization but will impart an additional momentum uncertainty. In other words, in the presence of decoherence the initial wave-packet expansion will proceed even more quickly compared to the decoherence-free case. The idea of observing the expansion of a wave-packet in order to test the predictions of quantum theory has been proposed already in the past, e.g., in~\cite{Collett2003a} for testing so-called continuous-spontaneous-localization models (CSL models) as alternatives to quantum theory~\cite{Ghirardi1990a}. Our approach takes a fresh look at this problem in the light of the novel possibilities provided by quantum optomechanics.

\begin{figure}[thbp]
 \begin{center}
  \includegraphics[width=0.49\linewidth]{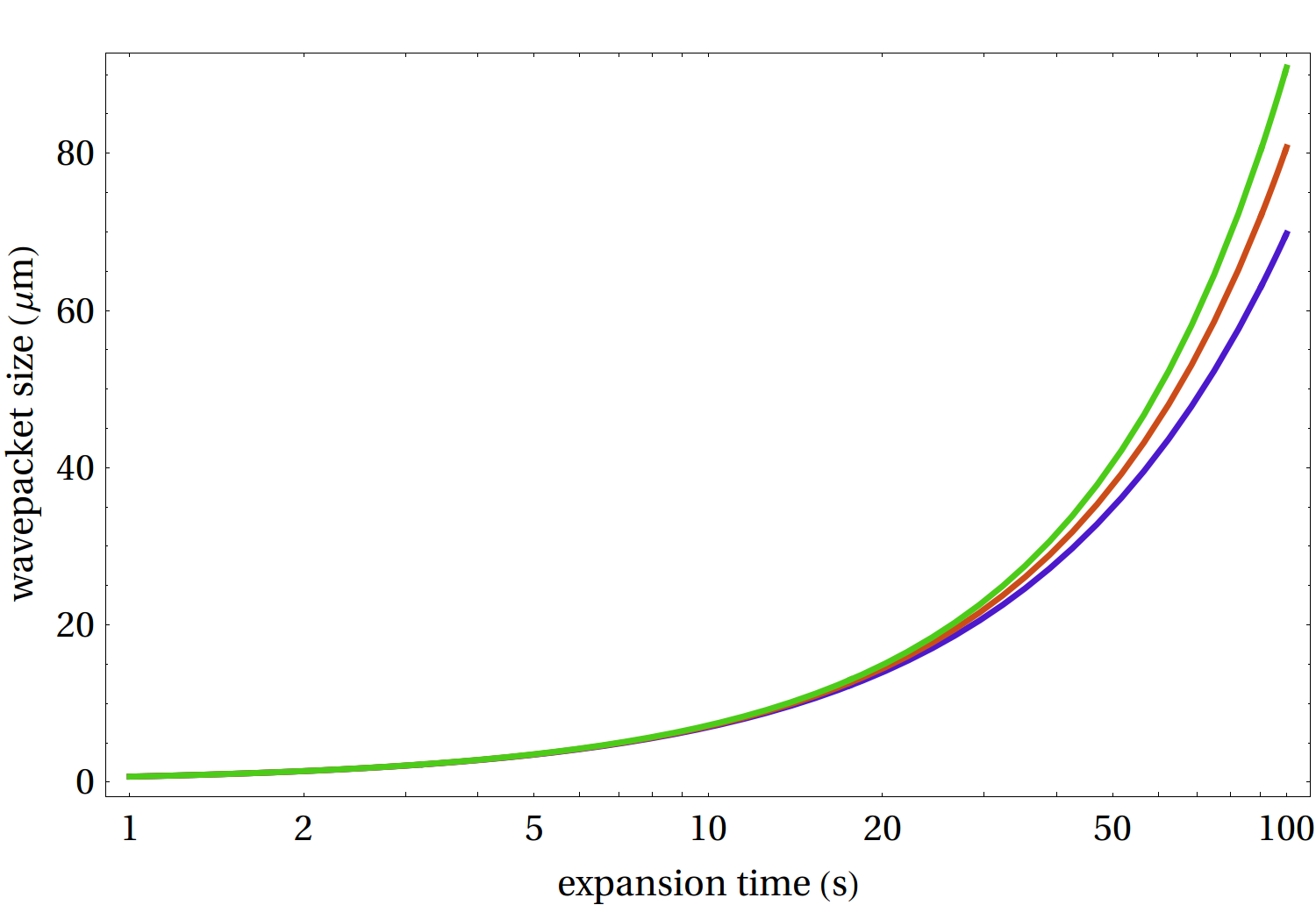}
  \includegraphics[width=0.49\linewidth]{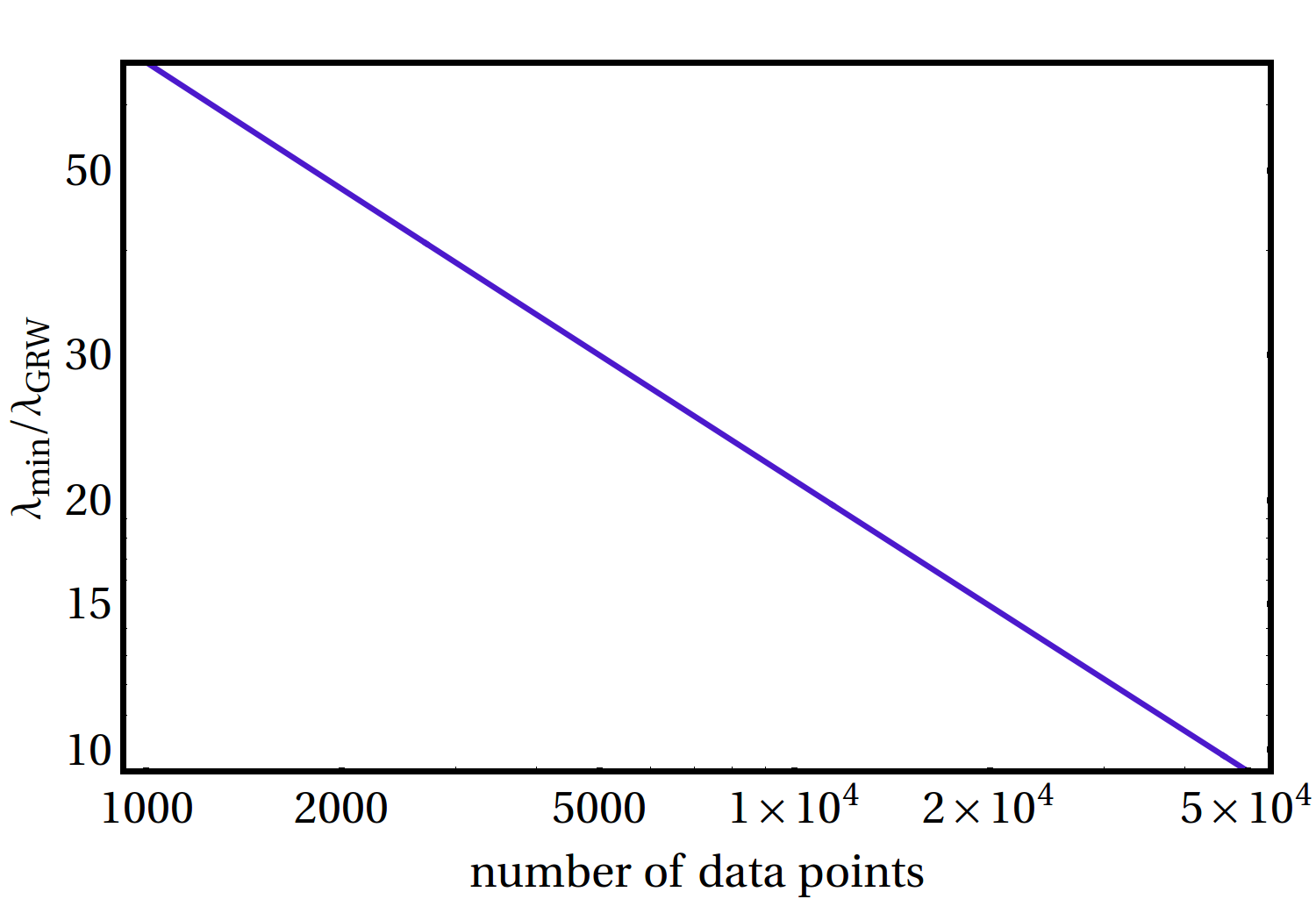}
  \caption{\textbf{(left)} Example for wave-packet expansion. (blue: no decoherence, red \& green increasing amount of decoherence). \textbf{(right)} Example for testing a modification of quantum theory (the CSL model). The number of data points measured determines the experimental error and therefore the minimum deviation from quantum theory we would notice experimentally. \label{fig::WAX:example}}
 \end{center}
\end{figure}

Figure \ref{fig::WAX:example}(left) illustrates the effect of this decoherence-induced acceleration of expansion of the wave-packet for the case of a fused-silica nanosphere with a radius of $120\,$nm. The blue line shows the expansion of the wave-packet if there is no decoherence. The red line shows the increased speed of wave-packet expansion in the presence of strong quantum decoherence due to the scattering, absorption and emission of blackbody radiation for $300\,$K environment temperature and $400\,$K internal temperature of the nanosphere. Decoherence due to gas collisions has been neglected for this example. The green line shows the expansion of the wave-packet as predicted by a macrorealistic modification of quantum theory (the CSL model for $a=100\,$nm and $\lambda=10^{-13}\,$Hz). The smaller the deviation from quantum theory, the stronger the requirements on the experimental accuracy needed to confirm such a deviation. The CSL model, for example, depends on two parameters. One of these parameters ($\lambda$) can be varied over a relatively wide range to fit experimental results. The smaller $\lambda$, the smaller is the deviation from quantum theory. In figure \ref{fig::WAX:example}(right), we plot the minimum $\lambda$ we could detect experimentally in units of the original value suggested, i.e., $\lambda_{GRW}=10^{-16}\,$Hz \cite{Ghirardi1986a}.

We discussed a possible implementation of an experiment to observe the \textbf{wa}ve-packet e\textbf{x}pansion of massive objects (WAX) in a recent study for ESA \cite{Kaltenbaek2012b}. A typical experimental run of WAX would consist of the following steps:
\begin{enumerate}
\setlength{\itemsep}{0pt}
\setlength{\parskip}{0pt}
\item	load a nanoparticle into an optical trap inside an optical cavity,
\item	move the particle to a predefined position along the cavity axis,
\item	cool the center-of-mass motion of the nanoparticle close to the quantum ground state,
\item	switch off the trap and let the particle's wavefunction expand freely for a time $t$,
\item	measure the position of the nanoparticle along the cavity axis,
\item	trap the particle again and repeat the steps above starting from step (2).
\end{enumerate}
By repeating this procedure often enough, one can determine the width of the wave packet for various values of $t$. Comparison with the predictions by quantum theory will then allow to detailed studying of possible sources of decoherence. Eventually, this analysis will even allow to place new experimental bounds on non-standard decoherence models, i.e., decoherence that is predicted to arise from various possible modifications of quantum theory (e.g., ``macrorealistic models'' \cite{Adler2009a}).

\begin{figure}[thbp]
 \begin{center}
  \includegraphics[width=0.49\linewidth]{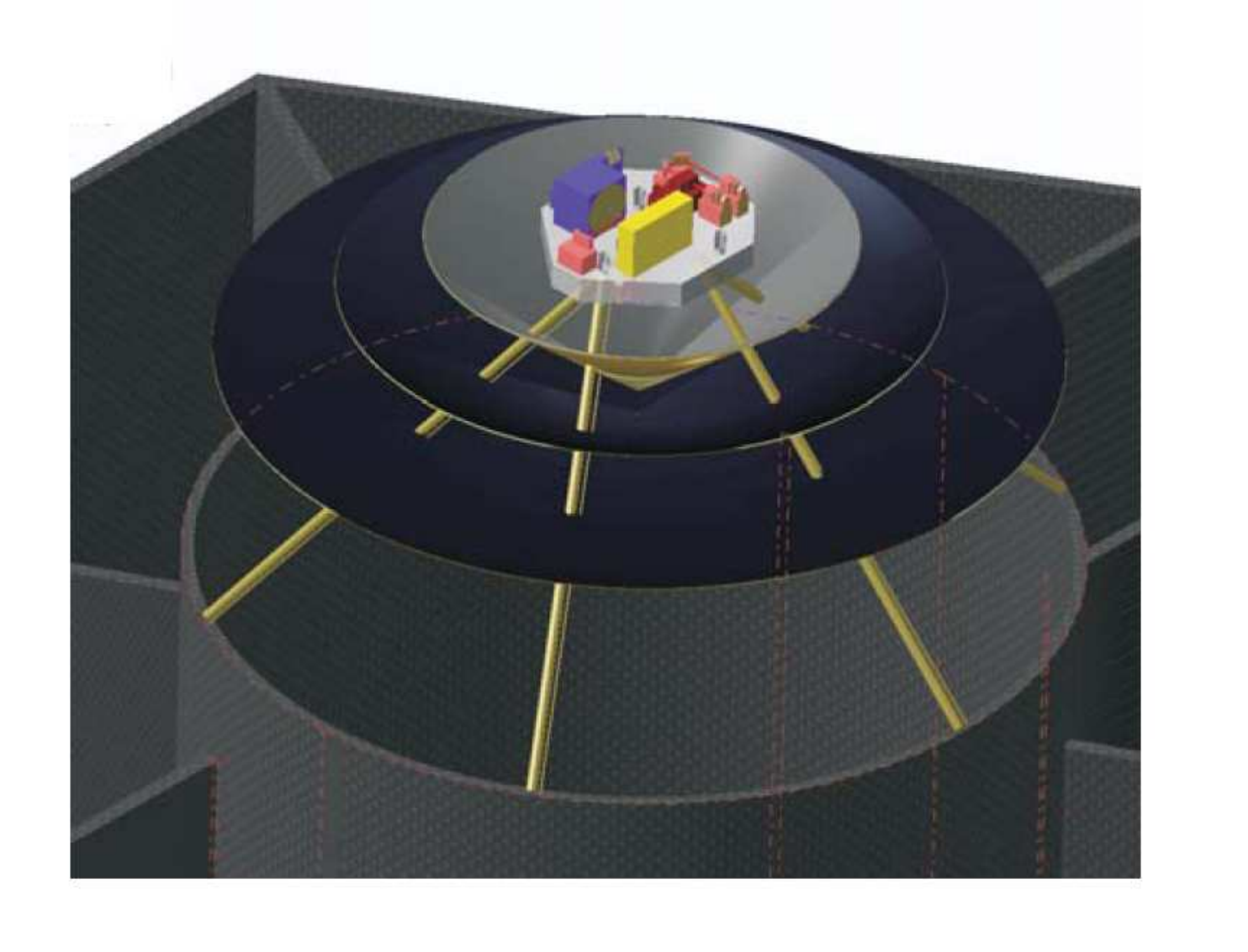}
  \includegraphics[width=0.49\linewidth]{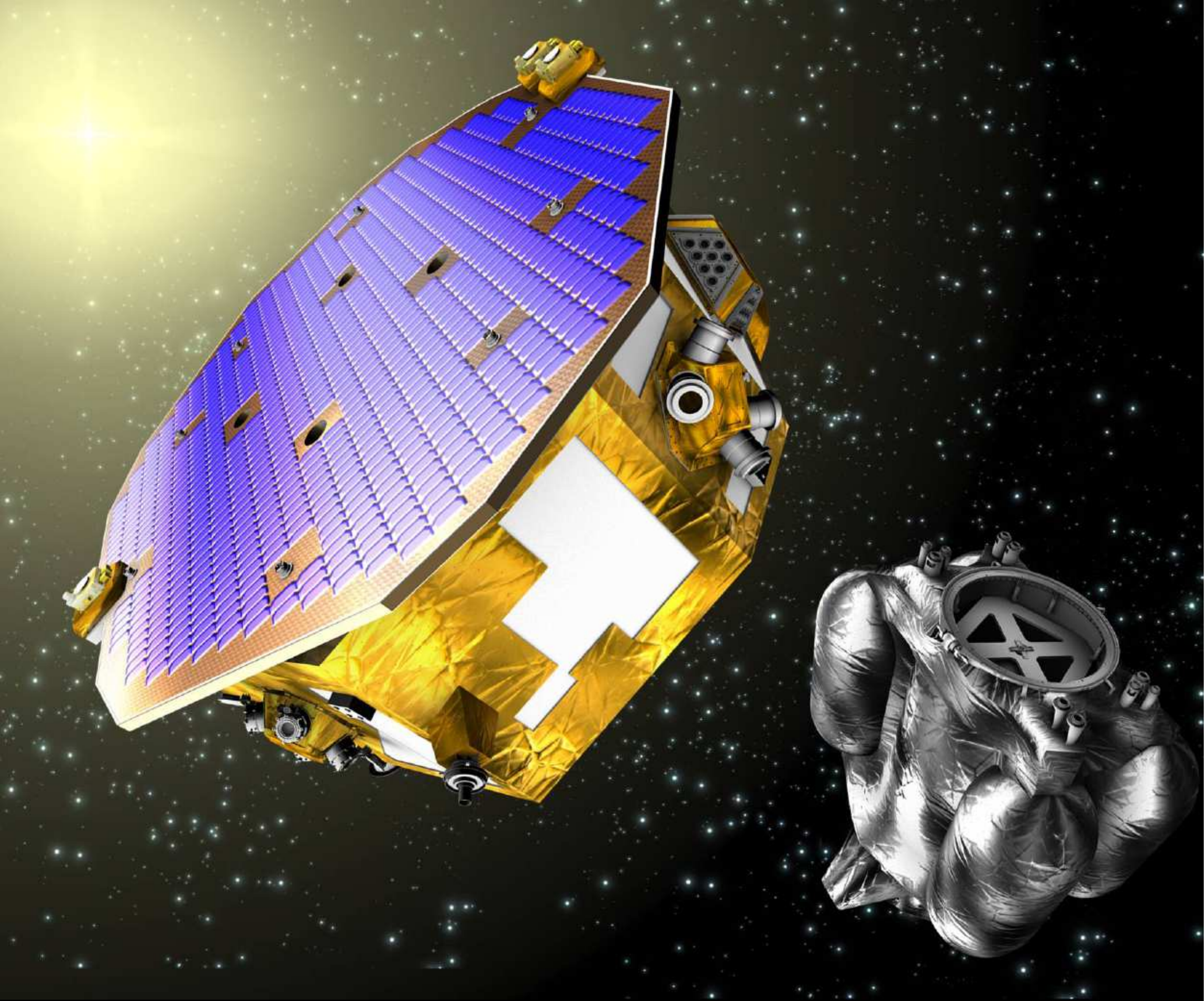}
  \caption{\textbf{(left)} Illustration of the thermal shield isolating the optical bench for WAX from the rest of the spacecraft~\cite{Kaltenbaek2012b}. \textbf{(right)} Artist's impression of LISA Pathfinder showing the solar array on top of the spacecraft and the separated propulsion module in the background. (Source: ESA \cite{LISAPF2008a}) \label{fig::WAX:spacecraft}}
 \end{center}
\end{figure}

Using WAX for performing tests of quantum theory beyond what has been achieved in experiments so far requires not only to go to significantly larger sizes but also to have long free-expansion times $t$ on the order of $10$--$100\,$s. Such free expansion times are hardly achievable on Earth because the path the particle would drop during that time would be between $500\,$m and $50\,$km. For this reason, we propose realizing WAX on a micro-gravity space platform. While this allows for long free-expansion times as needed for WAX, one also needs to keep standard quantum decoherence to a minimum (blackbody radiation, scattering of gas molecules). Achieving good vacuum and low temperatures on a spacecraft is, however, not trivial because the inside of spacecrafts is usually kept at room temperature in order for the electronic instruments aboard to work optimally. Because of the high temperature of the environment, components inside the spacecraft outgas at high rates leading to comparatively high pressures inside the spacecraft. In order to achieve good vacuum and low temperatures in space, one usually uses the same technology as on Earth, i.e., heavy vacuum and cryogenic equipment. The weight of that equipment leads to excessively high costs for space missions. We propose to achieve the vacuum and low-temperature ideal for WAX by performing the experiment on an optical bench that is thermally insulated from the rest of the spacecraft in order to facilitate thermal radiation as well as outgassing directly to space (see figure \ref{fig::WAX:spacecraft}(left) and references~\cite{Kaltenbaek2012a,Kaltenbaek2012b}). This should allow for environment temperatures of $30-40\,$K as well as ultra-high vacuum ($\lesssim 10^{-12}\,$Pa), which provides ideal conditions for WAX.

Under these conditions, WAX would allow coherent expansion of a nanosphere's wave-packet over times as large as $100\,$s and delocalization over distances comparable to the nanosphere's diameter. Because CSL localization is dominant in this regime, WAX would represent a stringent test for CSL (reducing the experimental limits on $\lambda$ by $\sim 6$ orders of magnitude compared to current tests).

Most of the optical equipment necessary for WAX (narrow-band lasers, optical components, stabilized interferometers, optical cavities etc.) is available or will soon be available for space applications. In particular, significant technological heritage exists in the form of the LISA Technology Package, which is the scientific instrument on LISA Pathfinder. For our studies, we assumed the spacecraft to be of the same type as LISA Pathfinder (see figure \ref{fig::WAX:spacecraft}(right) and~\cite{McNamara2008a}) in order to directly take advantage of that technological heritage.

We have analyzed the technical requirements for a space-based configuration of this type in detail in two recent studies ~\cite{Kaltenbaek2012a,Kaltenbaek2012b} and we refer the reader to these documents for more technical details. 

Finally, once the wave-packet has reached a certain coherent extension, one can also consider applying a physical interaction that will prepare the state of the nanosphere in an optomechanical cat state. According to quantum theory, a successful preparation of such a state will result in an interference pattern similar to the one occurring in the well-known double-slit experiment. Experiments of this type have been proposed both for ground-based experiments \cite{RomeroIsart2011b,RomeroIsart2011c} as well as for experiments in a space environment \cite{Kaltenbaek2012a,Kaltenbaek2012b}. In a space environment, such experiments could allow for testing quantum theory over a vastly larger parameter range than WAX \cite{Kaltenbaek2012a,Kaltenbaek2012b} -- hopefully resulting in increasingly ``burlesque'' situations of superpositions of macroscopically distinct states. Ultimately, one should attempt to increase the mass of the particle to a regime, where its own gravitational field can no longer be neglected. The views on what happens there are as diverse as they are on Schr\"odinger's cat\ldots

We note that this arXiv version of our paper is a simple reprint of the version published in early 2013 in Ref.~\cite{Kaltenbaek2013a} and submitted in late 2012. The only update in the bibliography was made in Ref.~\cite{Nimmrichter2013a}, which had been unpublished at the time.

\begin{acknowledgments}
We are grateful for discussions with Gerald Hechenblaikner and Ulrich Johann (Airbus Defense and Space) and with Keith Schwab (Caltech). We also thank Michelle Judd and the Keck Institute for Space Studies at Caltech for their support and hospitality, and the European Space Agency for support. R. K. acknowledges support from the Austrian Program for Advanced Research and Technology (APART) of the Austrian Academy of Sciences and support from the European Commission (Marie Curie, FP7-PEOPLE-2010-RG). M. A. acknowledges support by the European Space Agency and by the European Research Council (ERC StG QOM). 
\end{acknowledgments}

\bibliographystyle{apsrev}
\bibliography{schroedinger}

\end{document}